\documentstyle[12pt,epsfig,graphics]{article}
\begin{document}\bibliographystyle{plain}\begin{titlepage}
\renewcommand{\thefootnote}{\fnsymbol{footnote}}\hfill\begin{tabular}{l}
CUQM-94\\HEPHY-PUB 760/02\\UWThPh-2002-19\\hep-th/0210149\\October
2002\end{tabular}\\[1.5cm]\Large\begin{center}{\bf DISCRETE SPECTRA OF
SEMIRELATIVISTIC HAMILTONIANS}\\\vspace{0.8cm}\large{\bf Richard L.
HALL\footnote[3]{\normalsize\ {\em E-mail address\/}:
rhall@mathstat.concordia.ca}}\\[.3cm]\normalsize Department of Mathematics
and Statistics, Concordia University,\\1455 de Maisonneuve Boulevard West,
Montr\'eal, Qu\'ebec, Canada H3G 1M8\\[0.7cm]\large{\bf Wolfgang
LUCHA\footnote[1]{\normalsize\ {\em E-mail address\/}:
wolfgang.lucha@oeaw.ac.at}}\\[.3cm]\normalsize Institut f\"ur
Hochenergiephysik,\\\"Osterreichische Akademie der
Wissenschaften,\\Nikolsdorfergasse 18, A-1050 Wien,
Austria\\[0.7cm]\large{\bf Franz F.~SCH\"OBERL\footnote[2]{\normalsize\ {\em
E-mail address\/}: franz.schoeberl@univie.ac.at}}\\[.3cm]\normalsize Institut
f\"ur Theoretische Physik, Universit\"at Wien,\\Boltzmanngasse 5, A-1090
Wien, Austria\vfill {\normalsize\bf Abstract}\end{center}\normalsize We
review various attempts to localize the discrete spectra of semirelativistic
Hamiltonians of the form $H=\beta\sqrt{m^2+{\bf p}^2}+V(r)$ (defined, without
loss of generality but for definiteness, in three spatial dimensions) as
entering, for instance, in the spinless Salpeter equation;~every Hamiltonian
in this class of operators consists of the relativistic kinetic energy
$\beta\sqrt{m^2+{\bf p}^2},$ where $\beta>0$ allows for the possibility of
more than one particles of mass $m,$ and a spherically symmetric attractive
potential $V(r),$ $r\equiv|{\bf x}|.$ In general, accurate eigenvalues of a
nonlocal Hamiltonian operator can only be found by the use of a numerical
approximation procedure. Our main emphasis, however, is on the derivation of
rigorous semi-analytical expressions~for both upper and lower bounds to the
energy levels of such operators. We compare the~bounds obtained within
different approaches and present relationships existing between the bounds.

\vspace{3ex}

\renewcommand{\thefootnote}{\arabic{footnote}}\end{titlepage}

\normalsize

\section{Introduction: The ``spinless-Salpeter'' Hamiltonian}We investigate
the discrete spectrum---that is, the set of eigenvalues $E_{n\ell}$ below the
onset~of the essential spectrum, corresponding to bound states characterized
by the radial quantum number $n=1,2,3,\dots$ and the angular-momentum quantum
number $\ell=0,1,2,\dots$---of the semirelativistic ``spinless-Salpeter''
Hamiltonian for particles of mass $m$ and momentum~${\bf p},$\begin{equation}
H=\beta\sqrt{m^2+{\bf p}^2}+V(r)\ ,\quad r\equiv|{\bf x}|\ ,\quad\beta>0\
,\label{Eq:(sr)SSH}\end{equation}where $V(r)$ is an attractive central
potential in three spatial dimensions. As is evident~from the presence of the
relativistic kinetic-energy operator $\beta\sqrt{m^2+{\bf p}^2}$ (with
$\beta>0$ allowing, e.g., for more than one particle), this Hamiltonian may
be regarded as the generalization~of the nonrelativistic Schr\"odinger
Hamiltonian towards relativistic kinematics. The eigenvalue equation of the
operator (\ref{Eq:(sr)SSH}), called the ``spinless Salpeter equation,''
arises as a well-defined approximation to the Bethe--Salpeter formalism for
the description of bound states within~a relativistic quantum field theory.
Our main goal is to discuss bounds on the energy levels~$E.$

\section{Envelope theory for energy bounds}\label{Sec:ET}The envelope theory,
developed in Refs.~\cite{Hall83,Hall84}, constructs bounds on the discrete
eigenvalues $E_{n\ell}$ of a given operator $H$ by comparing the spectrum of
$H$ with the spectrum of a suitably formulated ``tangential problem,'' for
which some spectral information is known (or may~be obtained with
considerable more ease than for the original problem). More
specifically,~when applied to the (semirelativistic) Hamiltonian $H$ defined
in Eq.~(\ref{Eq:(sr)SSH}) \cite{Lucha00-HO,Lucha01-DMAI}, the envelope theory
compares $H$ with a corresponding ``tangential Hamiltonian''$$\widetilde
H=\beta\sqrt{m^2+{\bf p}^2}+vh(r)\ ,\quad v>0\ ,$$involving some ``basis
potential'' $h(r).$ Under the assumption that the interaction potential under
consideration, $V(r),$ is a smooth transformation $V(r)=g(h(r))$ of the basis
potential $h(r),$ where the function $g(h)$ has definite convexity, this
comparison yields rigorous bounds on the discrete eigenvalues $E_{n\ell}$ of
$H.$ More precisely, if the function $g(h)$ is convex, that~is, $g''(h)>0,$
we obtain lower energy bounds; if the function $g(h)$ is concave, that is,
$g''(h)<0,$ we obtain upper energy bounds. If the eigenvalues $\widetilde E$
of $\widetilde H$ are not known (analytically),~the envelope theory may take
advantage of the knowledge of suitable bounds on the eigenvalues, namely, of
lower bounds on $\widetilde E_{n\ell}$ for convex $g(h)$ or of upper bounds
on $\widetilde E_{n\ell}$ for concave~$g(h).$

Suppressing the quantum numbers $n,$ $\ell$ identifying the bound state under
consideration, we denote, for a given value of the coupling $v$ entering into
the ``tangential Hamiltonian''~$\widetilde H,$ the eigenvalues of $\widetilde
H$ or the appropriate---in the above sense---bounds to the latter by $e(v).$
The strict bounds on the discrete eigenvalues $E$ of $H$ predicted within the
framework of~the envelope technique may then be summarized in form of the
``principal envelope formula''~\cite{Lucha01-DMAI}\begin{equation}
E\approx\min_{v>0}[e(v)-ve'(v)+g(e'(v))]\ ,\label{Eq:PEF}\end{equation}where
the sign of approximate equality is used to recall that, for a definite
convexity~of~$g(h),$ the envelope theory yields lower bounds for convex
$g(h)$ and~upper bounds for concave $g(h).$

For the tangential problem posed by the (semirelativistic) ``tangential
Hamiltonian'' $\widetilde H,$ spectral information is, at present, available
if the basis potential $h(r)$ is either the Coulomb potential, that is,
$h(r)=-1/r,$ or the harmonic-oscillator potential, that is, $h(r)=r^2.$~The
energy bounds resulting in these two cases are discussed, in turn, in
Sec.~\ref{Sec:CLB} and Sec.~\ref{Sec:HOUB}. The case of the interaction
potential $V(r)$ in Eq.~(\ref{Eq:(sr)SSH}) being the harmonic-oscillator
potential~may be traced back to a nonrelativistic Schr\"odinger problem; this
is studied for its own in Sec.~\ref{Sec:HOP}.

\section{Harmonic-oscillator potential}\label{Sec:HOP}The ``relativistic
harmonic-oscillator problem,'' defined by the special case of $V(r)$
being~the harmonic-oscillator potential $V(r)=vr^2,$ may be analyzed by
observing \cite{Lucha96a,Lucha98O,Lucha98D,Lucha99Q,Lucha99A} that~in
momentum-space representation the semirelativistic Hamiltonian
$H=\beta\sqrt{m^2+{\bf p}^2}+vr^2$~of Eq.~(\ref{Eq:(sr)SSH}) simplifies to
the (nonrelativistic) Schr\"odinger operator ${\cal H}$ given
by\begin{equation}{\cal H}=-v\Delta+{\cal V}(r)\ ,\quad{\cal
V}(r)=\beta\sqrt{m^2+r^2}\ .\label{Eq:Ham-RHOP}\end{equation}Thus the
eigenvalue equation for $H$ reduces to an easier-to-treat nonrelativistic
Schr\"odinger problem with an effective interaction potential ${\cal V}(r)$
which is reminiscent of the square root.

\subsection{Upper bounds on harmonic-oscillator energy
levels}\label{Subsec:HOUB}The potential ${\cal V}$ is a concave transform of
a harmonic-oscillator potential. Thus we may~find upper bounds on the
eigenvalues $E_{n\ell}(v)$ of ${\cal H}$ (cf.\ Eq.~(2.2) of
Ref.~\cite{Lucha00-HO} or Eq.~(12) of
Ref.~\cite{Lucha01-DMAI}):\begin{equation}E_{n\ell}(v)\leq
\min_{r>0}\left[\beta\sqrt{m^2+\frac{P_{n\ell}^2(2)}{r^2}}+vr^2\right],\quad
P_{n\ell}(2)=2n+\ell-\frac{1}{2}\ .\label{Eq:HO-UBs}\end{equation}The
(dimensionless) numbers $P_{n\ell}(2)$ introduced here are related to the
eigenvalues ${\cal E}_{n\ell}$~of~the Hamiltonian $\widetilde H={\bf
p}^2+vr^2$ by ${\cal E}_{n\ell}=2\sqrt{v}P_{n\ell}(2).$

Upper bounds on the eigenvalues of self-adjoint operators bounded from below
may~also be found by combining the minimum--maximum principle
\cite{Reed78,Thirring90} with appropriate operator inequalities. (For
details, see, e.g., Sec.~3.4 of Ref.~\cite{Lucha96a}, Sec.~3.5 of
Ref.~\cite{Lucha98O}, Sec.~2.4 of Ref.~\cite{Lucha98D}, or Appendix~A of
Ref.~\cite{Lucha00-1DRCP}.) Exploiting the obvious positivity of
$(\sqrt{m^2+{\bf p}^2}-\mu)^2,$ where~$\mu$ is an arbitrary real parameter
(with the dimension of mass), we obtain a set of inequalities for the
square-root operator $\sqrt{m^2+{\bf p}^2}$ entering into the kinetic energy
of the Hamiltonian~(\ref{Eq:(sr)SSH}); cf.\ also Ref.~\cite{Martin88}:
$$\sqrt{m^2+{\bf p}^2}\le\frac{{\bf p}^2+m^2+\mu^2}{2\mu}\quad\forall\ \mu>0\
.$$(This operator inequality may also be found by constructing the tangent
line $a({\bf p}_1^2){\bf p}^2+b({\bf p}_1^2)$ to the square root
$\sqrt{m^2+{\bf p}^2}$ at the point of contact ${\bf p}_1^2=\mu^2-m^2;$ see
Sec.~4 of Ref.~\cite{Lucha01-DMAI}.) Amending this result by the interaction
potential $V(r),$ one finds that the spinless-Salpeter Hamiltonian
(\ref{Eq:(sr)SSH}) satisfies$$H\le H_{\rm NR}(\mu)\equiv\beta\frac{{\bf
p}^2+m^2+\mu^2}{2\mu}+V(r)\quad\forall\ \mu>0\ .$$The min--max principle then
tells us that the discrete eigenvalues $E_{n\ell}$ of the semirelativistic
Hamiltonian $H$ of Eq.~(\ref{Eq:(sr)SSH}) are bounded from above by the
corresponding discrete eigenvalues $E_{{\rm NR},n\ell}(\mu)$ of the
nonrelativistic Hamiltonian $H_{\rm NR}(\mu)$ and consequently also by the
minimum $\bar E_{n\ell}$ with respect to $\mu$ of all these upper bounds
\cite{Lucha96a}:$$E_{n\ell}\le\bar E_{n\ell}\equiv\min_{\mu>0}E_{{\rm
NR},n\ell}(\mu)\ .$$Applying these general findings to the case of the
harmonic-oscillator potential, one ends~up with the following upper bounds
for all energy levels of the relativistic harmonic-oscillator problem (see
Appendix~A of Ref.~\cite{Lucha99A}; for the ease of comparison, we express
here the bounds of Ref.~\cite{Lucha99A} in terms of $P_{n\ell}(2)$ rather
than in terms of $E_{{\rm NR},n\ell}(\mu)$):
\begin{equation}E_{n\ell}(v)\le\bar E_{n\ell}\equiv\min_{\mu>0}\left[
\beta\frac{m^2+\mu^2}{2\mu}+\sqrt{\frac{2\beta v}{\mu}}P_{n\ell}(2)\right],
\quad P_{n\ell}(2)=2n+\ell-\frac{1}{2}\ .\label{Eq:HO-MMOI-UBs}\end{equation}

\newpage Despite their seemingly different form, the upper bounds in
Eqs.~(\ref{Eq:HO-UBs}) and (\ref{Eq:HO-MMOI-UBs}) are, in~fact, identical, in
the sense that the two functions on the right-hand side of these
inequalities~have their minimum at the same critical point and take the same
miminal value at this point.

In order to demonstrate this equivalence, let us start from the ``min--max''
expression~(\ref{Eq:HO-MMOI-UBs}) for these upper bounds. For simplicity of
notation, we suppress for the moment the quantum numbers $n,$ $\ell$ which
identify the bound states under consideration. The upper bound $\bar E$~then
reads$$\bar E=\min_{\mu>0}
\left[\beta\frac{m^2+\mu^2}{2\mu}+\sqrt{\frac{\beta v}{2\mu}}2P\right].$$We
introduce the abbreviation $q=p^2,$ define the function $f(q)=\sqrt{m^2+q},$
which~satisfies$$f'(q)=\frac{1}{2f(q)}\ ,\quad f^2(q)=m^2+q\ ,$$and identify
the (positive but otherwise arbitrary) parameter $\mu$ according to
$\mu=f(q).$~Next, we introduce a function $F$ by the definition
$F(v)=2P\sqrt{v}.$ A couple of algebraic steps then recasts the upper bound
$\bar E$ into the form$$\bar E=\min_{q>0}[F(\beta vf')+\beta(f-qf')]\ .$$The
critical points of this latter expression are obtained by differentiation
with respect~to~$q.$ Since $f''(q)\ne 0,$ this yields$$\hat q=vF'(\beta
vf'(\hat q))$$and$$\bar E=F(\beta vf'(\hat q))-\beta vf'(\hat q)F'(\beta
vf'(\hat q))+\beta f(\hat q)\ .$$

Let us define a new problem, related to the above one by a Legendre
transformation~\cite {Gelfand}, which has the same minimum and the same
critical point, but the function to be minimized is {\em different}. Let
$u(q)=\beta vf'(q)>0$ and consider the energy expression$$\hat
E=\min_{u>0}[F(u)-uF'(u)+\beta f(vF'(u))]\ .$$Since $F''(u)\ne 0,$ we find
that the critical point is given by $\hat u=\beta vf'(vF'(\hat u)).$ But,
given the definition of $u,$ this equation is equivalent to the earlier
critical equation $\hat q=vF'(\beta vf'(\hat q)).$ Meanwhile, the critical
{\em value\/} $\hat E$ is also the same as the critical value $\bar E$
because we have$$\hat E=F(\beta vf'(\hat q))-\beta vf'(\hat q)F'(\beta
vf'(\hat q))+\beta f(\hat q)=\bar E\ .$$

Having shown that $\bar E$ yields the same results as $\hat E,$ we now
transform $\hat E$ into the form~(\ref{Eq:HO-UBs}). First of all, we note
that $F(u)-uF'(u)=P\sqrt{u}.$ Next, we define a new dummy~minimization
variable $r>0$ by$$\sqrt{u}=\frac{vr^2}{P}\ .$$From this, it follows
that$$F'(u)=\frac{P}{\sqrt{u}}=\frac{P^2}{vr^2}\ .$$Finally, remembering the
definition $f(q)=\sqrt{m^2+q}$ of $f(q),$ we may transform $\hat E$ into the
expression on the right-hand side of the inequality (\ref{Eq:HO-UBs}); this
establishes the equivalence~of~the upper bounds (\ref{Eq:HO-UBs}) and
(\ref{Eq:HO-MMOI-UBs}).

\subsection{Lower bounds on harmonic-oscillator energy
levels}\label{Subsec:HOLB}The potential ${\cal V}$ in Eq.~(\ref{Eq:Ham-RHOP})
is a convex transform of a linear potential. Therefore we may~find lower
bounds on the eigenvalues $E_{n\ell}(v)$ of ${\cal H}$ (cf.\ Eq.~(2.2) of
Ref.~\cite{Lucha00-HO} or Eq.~(12) of
Ref.~\cite{Lucha01-DMAI}):\begin{equation}E_{n\ell}(v)\ge
\min_{r>0}\left[\beta\sqrt{m^2+\frac{P_{n\ell}^2(1)}{r^2}}+vr^2\right],
\label{Eq:HO-LBs}\end{equation}where the (dimensionless) numbers
$P_{n\ell}(1)$ introduced here are related to the eigenvalues~${\cal
E}_{n\ell}$ of the Hamiltonian $\widetilde H={\bf p}^2+vr$ by$${\cal
E}_{n\ell}=3\left[\frac{vP_{n\ell}(1)}{2}\right]^{2/3}\ ;$$for the lowest
states ($n=1,2,\dots,5,$ $\ell=0,1,\dots,5$) these numbers may be found
in~Table~\ref{Tab:P(1)-numbers}.

\begin{table}[ht]\caption{Numerical values of the energy-related numbers
$P_{n\ell}(1)$ for the lowest energy levels.}\label{Tab:P(1)-numbers}
\begin{center}\begin{tabular}{ccr}\hline\hline&&\\[-1.5ex]
\multicolumn{1}{c}{$n$}&\multicolumn{1}{c}{$\ell$}&
\multicolumn{1}{c}{$P_{n\ell}(1)$}\\[1ex]\hline\\[-1.5ex]
1&0&1.37608\\2&0&3.18131\\3&0&4.99255\\4&0&6.80514\\5&0&8.61823\\[.5ex]
1&1&2.37192\\2&1&4.15501\\3&1&5.95300\\4&1&7.75701\\5&1&9.56408\\[1ex]
\hline\hline\end{tabular}$\qquad$\begin{tabular}{ccr}\hline\hline&&\\[-1.5ex]
\multicolumn{1}{c}{$n$}&\multicolumn{1}{c}{$\ell$}&
\multicolumn{1}{c}{$P_{n\ell}(1)$}\\[1ex]\hline\\[-1.5ex]
1&2&3.37018\\2&2&5.14135\\3&2&6.92911\\4&2&8.72515\\5&2&10.52596\\[.5ex]
1&3&4.36923\\2&3&6.13298\\3&3&7.91304\\4&3&9.70236\\5&3&11.49748\\[1ex]
\hline\hline\end{tabular}$\qquad$\begin{tabular}{ccr}\hline\hline&&\\[-1.5ex]
\multicolumn{1}{c}{$n$}&\multicolumn{1}{c}{$\ell$}&
\multicolumn{1}{c}{$P_{n\ell}(1)$}\\[1ex]\hline\\[-1.5ex]
1&4&5.36863\\2&4&7.12732\\3&4&8.90148\\4&4&10.68521\\5&4&12.47532\\[.5ex]
1&5&6.36822\\2&5&8.12324\\3&5&9.89276\\4&5&11.67183\\5&5&13.45756\\[1ex]
\hline\hline\end{tabular}\end{center}\end{table}

Lower bounds on the eigenvalues of (self-adjoint) Hamiltonians involving the
relativistic kinetic energy $\sqrt{m^2+{\bf p}^2}$ may also be found by
taking into account the obvious positivity~of $[{\bf
p}^2\xi^2-m^2(1-\xi^2)]^2$ for an arbitrary real parameter $\xi,$ which may
be converted into the~set of operator inequalities\begin{equation}
\sqrt{m^2+{\bf p}^2}\ge|{\bf p}|\sqrt{1-\xi^2}+m\xi\ ,\quad 0\le\xi\le 1\
.\label{Eq:RKE-LB}\end{equation}Applying these general findings to the
semirelativistic Hamiltonian $H$ of Eq.~(\ref{Eq:(sr)SSH}), with $V(r)$ being
the harmonic-oscillator potential, we obtain {\em lower\/} bounds on {\em
all\/} discrete energy levels $E_{n\ell}$ of the relativistic
harmonic-oscillator problem (for details, see Appendix~A of
Ref.~\cite{Lucha99A}):\begin{equation}E_{n\ell}(v)\ge\underline
E_{n\ell}\equiv\max_{0\le\xi\le 1}[\beta\xi
m+\beta^{2/3}(1-\xi^2)^{1/3}E_{n\ell}(v;m=0)]\
,\label{Eq:HO-xi-LBs}\end{equation}where $E_{n\ell}(v;m=0)$ denote the
corresponding discrete eigenvalues of the operator $|{\bf p}|+vr^2,$ i.e., of
the Hamiltonian (\ref{Eq:(sr)SSH}) for $\beta=1$ and a vanishing particle
mass $m.$ For states with~$\ell=0,$ for instance, these ``zero-mass'' energy
levels are given, in terms of the zeros $z_n$ of the Airy function ${\rm
Ai}(z)$ \cite{Abramowitz} ($-z_n=2.33810,\,4.08794,\,5.52055,\,\dots$), by
$E_{n0}(v;m=0)=-v^{1/3}z_n$~\cite{Lucha99A}.

As in Sec.~\ref{Subsec:HOUB} we observe that the lower bounds in
Eqs.~(\ref{Eq:HO-LBs}) and (\ref{Eq:HO-xi-LBs}) appear, at first~sight, to be
rather different. A closer inspection, however, reveals that these bounds are
identical.

In order to present an analytical proof of the equivalence of the lower
bounds (\ref{Eq:HO-LBs}) and~(\ref{Eq:HO-xi-LBs}) on the energy eigenvalues
of the relativistic harmonic-oscillator problem, we recall that the envelope
lower bounds in (\ref{Eq:HO-LBs}) are based (in the momentum-space
representation~of~$H$)~on~the observation that, since its second derivative
with respect to $|{\bf p}|$ is positive, the square-root operator
$\sqrt{m^2+{\bf p}^2}$ of the relativistic kinetic energy is bounded from
below by tangent~lines $a(t)+b(t)|{\bf p}|;$ this may be expressed by the
geometrical inequality $\sqrt{m^2+{\bf p}^2}\ge a(t)+b(t)|{\bf p}|,$ where,
by elementary calculus,$$a(t)=\frac{m^2}{\sqrt{m^2+t^2}}\ ,\quad
b(t)=\frac{t}{\sqrt{m^2+t^2}}\ ,$$and $|{\bf p}|=t$ is the point of contact
between the square root of the relativistic kinetic energy and its
straight-line lower approximation. The equivalence of the above two
expressions for the harmonic-oscillator lower bound is then shown by a simple
change of variable from~$t$~to~$\xi$ given explicitly by$$\xi=\frac{a(t)}{m}\
,$$which, after a little algebra, leads to $b(t)=\sqrt{1-\xi^2}.$ For these
coefficient functions $a(t)$~and $b(t)$ the above geometrical inequality
becomes exactly the lower bound (\ref{Eq:RKE-LB}) on the relativistic kinetic
energy. Consequently, with the basic inequalities being identical, the
resulting lower energy bounds have to be identical too.

\section{Convex transform of Coulomb potential}\label{Sec:CLB}For the
``relativistic Coulomb problem'' posed by the spinless-Salpeter
Hamiltonian~(\ref{Eq:(sr)SSH})~with $V(r)$ being the Coulomb potential
$V(r)=-v/r,$ for coupling constants $v$ smaller than $\beta v_{\rm c},$ that
is, $v<\beta v_{\rm c},$ where the critical value $v_{\rm c}$ of the Coulomb
coupling constant $v$ is given by$$v_{\rm c}=\frac{2}{\pi}\ ,$$a lower bound
to the ground-state energy eigenvalue $E_0$ of $H$ (the bottom of the
spectrum of $H$) has been derived by Herbst \cite{Herbst77}:\begin{equation}
E_0\ge\beta m\sqrt{1-\left(\frac{\sigma v}{\beta}\right)^{2}}\
,\quad\sigma\equiv\frac{\pi}{2}\ .\label{Eq:RCP-HLB}\end{equation}For some
part of the allowed range of the coupling $v,$ this lower bound has been
improved~by Martin and Roy \cite{Martin89}:\begin{equation}E_0\ge\beta
m\sqrt{\frac{1}{2}\left[1+\sqrt{1-\left(\frac{2v}{\beta}\right)^2}\right]}\
,\quad v<\frac{\beta}{2}\ .\label{Eq:RCP-MRLB}\end{equation}In
Ref.~\cite{Lucha01-DMAI}, we recast this Martin--Roy bound into a class of
lower bounds on the spectrum~of $H$ all of which are of the form of Herbst's
bound (\ref{Eq:RCP-HLB}) [but, of course, slightly weaker than~the
Martin--Roy bound (\ref{Eq:RCP-MRLB})]; the new lower bounds are parametrized
by a real parameter~$\sigma\ge 1$:\begin{equation}E_0\ge\beta
m\sqrt{1-\left(\frac{\sigma v}{\beta}\right)^2}\ ,\quad
v\le\beta\frac{\sqrt{\sigma^2-1}}{\sigma^2}<\frac{\beta}{2}\
.\label{Eq:RCP-NLB}\end{equation}The constraint on the Coulomb coupling $v$
in Eq.~(\ref{Eq:RCP-NLB}) automatically guarantees $v\le\beta/\sigma;$ this
latter constraint is a necessary and sufficient condition for the reality of
the lower~energy bounds in Eq.~(\ref{Eq:RCP-NLB}).

From the lower bounds (\ref{Eq:RCP-HLB}) or (\ref{Eq:RCP-NLB}) on the
spectrum of the relativistic Coulomb problem, for the class of potentials
$V(r)$ which are convex transforms $V(r)=g(h(r)),$ $g''>0,$~of~the Coulomb
potential $h(r)=-1/r$ an (``absolute'') envelope lower bound on the
ground-state eigenvalue $E_0$ (and, hence, on the entire spectrum) of the
spinless-Salpeter Hamiltonian~(\ref{Eq:(sr)SSH}) may be found (cf.\ Eq.~(16)
of Ref.~\cite{Lucha01-DMAI}):\begin{equation}
E_0\ge\min_{r>0}\left[\beta\sqrt{m^2+\frac{P^2}{r^2}}+V(r)\right],\quad
v<\beta v_P\ .\label{Eq:SSE-LB}\end{equation}Here we introduced a parameter
$P$ by defining $P=1/\sigma.$ Accordingly, the boundary value~$v_P$ of the
Coulomb coupling constant $v$ is given, when arising from demanding the
Hamiltonian (\ref{Eq:(sr)SSH}) to be bounded from below, by the critical
coupling $v_{\rm c},$ i.e., $v_P=v_{\rm c},$ or, when arising~from the region
of validity of our new family of Coulomb lower bounds (\ref{Eq:RCP-NLB}),
by\begin{equation}v_P=P\sqrt{1-P^2}<\frac{1}{2}\
.\label{Eq:BV}\end{equation}It is easy to convince oneself that the Coulomb
lower energy bound on the right-hand~side~of the inequality (\ref{Eq:SSE-LB})
is a monotone increasing function $\underline E(P)$ of the parameter $P.$ To
this~end, we rewrite the inequality (\ref{Eq:SSE-LB}) in the
form$$E_0\ge\underline E(P)=\min_{r>0}F(r,P)\
,$$with$$F(r,P)\equiv\beta\sqrt{m^2+\frac{P^2}{r^2}}+V(r)\ .$$Remembering
that $V(r)$ is the transform $V(r)=g(h(r))$ of $h(r)=-1/r,$ and assuming~that
$g'(h)>0,$ the critical point $\hat r(P)$ (corresponding to the minimum of
$F(r,P)$ for a given $P$) may be found, from$$\left.\frac{\partial
F(r,P)}{\partial r}\right|_{r=\hat r(P)}=0\ ,$$by solving$$\frac{\beta
P^2}{\sqrt{m^2\hat r^2+P^2}}=g'(h(\hat r))\ .$$The substitution of the
solution $\hat{r}(P)$ of the latter equation in $F(r,P)$ gives the
lower~bound $\underline E(P)=F(\hat r(P),P);$ here we are interested in the
total derivative of $\underline E(P)$ with respect~to~$P.$ Formally, this
derivative would be given by$$\frac{{\rm d}\underline E(P)}{{\rm
d}P}=\frac{\partial F(\hat r(P),P)}{\partial P}+\left.\frac{\partial
F(r,P)}{\partial r}\right|_{r=\hat r(P)}\frac{{\rm d}\hat r(P)}{{\rm d}P}\
.$$At the critical point, however, the first factor of the second term
vanishes and one~is~left~with$$\frac{{\rm d}\underline E(P)}{{\rm
d}P}=\frac{\partial F(\hat r(P),P)}{\partial P}=\frac{\beta P}{\hat
r\sqrt{m^2\hat r^2+P^2}}>0\ .$$From this we conclude that the lower bound
$\underline E(P)$ is a monotone increasing function of $P.$

The existence of the upper bound $v<\beta v_P<\beta/2$ on the Coulomb
coupling constant~$v$~is reflected in a corresponding constraint on the
parameters of the interaction potential~$V(r).$ The envelope theory
constructs a ``tangential potential'' to $V(r)$ at a point of contact~$r=t.$
For a potential $V(r)$ which is a convex transform $V(r)=g(h(r))$ of the
basis potential~$h(r),$ this tangential potential forms a lower bound to
$V(r);$ for $h(r)$ being the Coulomb potential, $h(r)=-1/r,$ this lower bound
is a Coulomb potential with the effective coupling constant $v(t)=t^2V'(t),$
shifted by the constant $V(t)+tV'(t)$:$$V(r)\ge\widetilde
V(r)\equiv-\frac{t^2V'(t)}{r}+V(t)+tV'(t)\ .$$The relativistic Coulomb
problem is posed by the ``Coulombic'' Hamiltonian$$H_{\rm
C}(v)=\beta\sqrt{m^2+{\bf p}^2}-\frac{v}{r}\ ,\quad v<\beta v_{\rm c}\
,$$which is nothing else but the tangential Hamiltonian $\widetilde H$ with
$h(r)=-1/r.$ In the notation~of Sec.~\ref{Sec:ET}, the eigenvalues $E_{\rm
C}(v)$ of $H_{\rm C}(v)$ are bounded from below according to $E_{\rm C}(v)\ge
e(v),$ with $e(v)$ given by Herbst's bound (\ref{Eq:RCP-HLB}), our new bound
(\ref{Eq:RCP-NLB}), or the Martin--Roy~bound~(\ref{Eq:RCP-MRLB}). Expressed
in terms of the Coulombic Hamiltonian $H_{\rm C}(v),$ the semirelativistic
Hamiltonian (\ref{Eq:(sr)SSH}) thus satisfies the inequality $H\ge H_{\rm
C}(v(t))+V(t)+tV'(t);$ its eigenvalues $E$ are bounded from below according
to$$E_0\ge E_{{\rm C},0}(v(t))+V(t)+tV'(t)\ge e(v(t))+V(t)+tV'(t)\ .$$The
optimized lower bound on the discrete spectrum of $H$ is then found by
maximizing~the expression on the right-hand side of this latter inequality
with respect to the point of contact $r=t.$ The equation which determines the
critical point $\hat t$ has to be solved~by~observing~that, at the critical
point $r=\hat t,$ the effective Coulomb coupling $v(t)$ still has to
satisfy~$v(\hat t)<\beta v_P;$ this nontrivial requirement imposes the
announced ``Coulomb coupling constant constraint'' on the coupling parameters
entering in the potential~$V(r).$

It goes without saying that, when applying our ``Herbst-based'' lower energy
bounds of Eq.~(\ref{Eq:RCP-NLB}), care has to be taken in order to assure
that the chosen values of the parameters~in the potential $V(r)$ under
consideration respect this Coulomb coupling constant constraint.

Let us illustrate these general considerations by applying them to a
Coulomb-plus-linear or (in view of its shape) ``funnel'' potential
\begin{equation}V(r)=-\frac{c_1}{r}+c_2r\ ,\quad c_1\ge 0\ ,\quad c_2\ge 0\
.\label{Eq:FP}\end{equation}In this case, the effective Coulomb coupling
$v(t)$ explicitly reads $v(t)=c_1+c_2t^2$ and, for~the set of ``Herbst-like''
lower energy bounds $e(v)$ given by Eqs.~(\ref{Eq:RCP-HLB}) and
(\ref{Eq:RCP-NLB}), the critical point~$\hat t$ is, after some algebra,
determined by the relation$$v(\hat t)=\frac{\beta P^2}{\sqrt{m^2\hat
t^2+P^2}}\ .$$The above Coulomb coupling constant constraint becomes an upper
bound on a particular linear combination of the two involved coupling
parameters $c_1$ and $c_2$ (cf.\ Sec.~6 of Ref.~\cite{Lucha01-DMAI}):
\begin{equation}c_1+\frac{P^2}{m^2}\left(\frac{P^2}{v_P^2}-1\right)c_2<\beta
v_P\ .\label{Eq:CCCC}\end{equation}

The Coulomb lower bounds (\ref{Eq:SSE-LB}) on the spectrum of the
semirelativistic Hamiltonian~(\ref{Eq:(sr)SSH}) may be optimized by seeking
that value of the parameter $P=1/\sigma$ which, for given values~of the mass
$m$ of the bound-state constituent(s) and of the coupling parameters $c_1$
and $c_2$ in~the Coulomb-plus-linear potential (\ref{Eq:FP}), solves the
Coulomb coupling constant constraint~(\ref{Eq:CCCC}). Taking into account the
definition (\ref{Eq:BV}) of the boundary value $v_P$ of the Coulomb
coupling~$v,$ this optimum value of $P$ is obtained as the solution of the
relation\begin{equation}\frac{c_2\sin^4t}{\cos^2t(\beta\sin t\cos
t-c_1)}=m^2\ ,\quad P\equiv\sin t\ .\label{Eq:P(m)}\end{equation}

The set (\ref{Eq:SSE-LB}) of envelope lower bounds distinguished from each
other by the parameter~$P$ may be improved somewhat by taking into account
the Martin--Roy lower bound (\ref{Eq:RCP-MRLB})~on~the energy spectrum of the
relativistic Coulomb problem. To this end, let us denote the energy bound on
the right-hand side of the inequality~(\ref{Eq:RCP-MRLB}) by
$e(v)$:$$e(v)=\beta
m\sqrt{\frac{1}{2}\left[1+\sqrt{1-\left(\frac{2v}{\beta}\right)^2}\right]}\
.$$Performing the change of variables$$v=\frac{\beta}{2}\sin 2t\ ,\quad 0\le
t<\frac{\pi}{4}\ ,$$we find$$e(v)=\beta m\cos t\ ,\quad e(v)-ve'(v)=\beta
m\frac{\cos^3t}{\cos 2t}\ ,$$while from $e'(v)=h(r)$ we get$$r=\frac{\cos
2t}{m\sin t}\ .$$Upon inserting these intermediate results in our principal
envelope formula (\ref{Eq:PEF}), we arrive~at the ``Martin--Roy-based''
Coulomb lower bound$$E_0\ge\min_{0\le t<\pi/4}\left[\beta
m\frac{\cos^3t}{\cos 2t}+V\left(\frac{\cos 2t}{m\sin t}\right)\right].$$

\section{Concave transform of harmonic-oscillator potential}\label{Sec:HOUB}
For the class of potentials $V(r)$ which are concave transforms
$V(r)=g(h(r)),$ $g''<0,$~of~the harmonic-oscillator potential $h(r)=r^2$
envelope upper bounds on the eigenvalues $E_{n\ell}$~of~the spinless-Salpeter
Hamiltonian (\ref{Eq:(sr)SSH}) may be found for all energy levels (cf.\
Eq.~(18) of Ref.~\cite{Lucha01-DMAI}):\begin{equation}
E_{n\ell}\le\min_{r>0}\left[\beta\sqrt{m^2+\frac{P^2}{r^2}}+V(r)\right],\quad
P\equiv P_{n\ell}(2)=2n+\ell-\frac{1}{2}\ .\label{Eq:SSE-HOUB}\end{equation}

\section{Concave transform of linear potential}\label{Sec:LUB}For the class
of potentials $V(r)$ which are concave transforms $V(r)=g(h(r)),$
$g''<0,$~of~the linear potential $h(r)=r$ the ``harmonic-oscillator based''
envelope upper bounds (\ref{Eq:SSE-HOUB}) on~the eigenvalues $E_{n\ell}$ of
the spinless-Salpeter Hamiltonian (\ref{Eq:(sr)SSH}) may be (significantly)
improved by considering the composition of the envelope approximations for
the kinetic-energy operator and the interaction-potential operator along the
following lines. In the present investigation we focus our attention to
attractive potentials $V(r),$ that is, to all potentials which exhibit~a
monotone increase with increasing radial coordinate $r$: $g'>0.$ Every
function $g(r)$ that~is~a monotone increasing [$g'(r)>0$] and concave
[$g''(r)<0$] function of $r$ is a concave function $f(r^2)$ of $r^2$:
$g(r)=f(r^2)$ with $f''(r^2)<0.$ Consequently, all potentials $V(r)$ studied
in this section also belong to the class of potentials considered in
Sec.~\ref{Sec:HOUB}.

The square-root operator $k({\bf p}^2)\equiv\sqrt{m^2+{\bf p}^2}$ of the
relativistic kinetic energy, regarded as a function of ${\bf p}^2,$ is
monotone increasing (i.e., $k'({\bf p}^2)>0$) and concave (i.e., $k''({\bf
p}^2)<0$). Let us assume that the interaction potential $V(r)$ exhibits a
similar behaviour, that is,~that the operator $V(r)=g(r)$ is monotone
increasing, i.e., $g'(r)>0,$ and concave, i.e., $g''(r)<0.$ These operators
are bounded from above by their respective upper-tangent approximations:
\begin{eqnarray*}&&k({\bf p}^2)\le a_1(q)+b_1(q){\bf p}^2\ ,\\[1ex]&&g(r)\le
a_2(t)+b_2(t)r\ ,\end{eqnarray*}with\begin{eqnarray}&&b_1(q)=k'(q)\ ,\quad
a_1(q)=k(q)-qk'(q)\ ,\nonumber\\[1ex]&&b_2(t)=g'(t)\ ,\quad
a_2(t)=g(t)-tg'(t)\ ,\label{Eq:comp-coeff}\end{eqnarray}where ${\bf p}^2=q$
denotes the point of contact between the kinetic-energy operator $k({\bf
p}^2)$ and~its upper-tangent approximation $a_1(q)+b_1(q){\bf p}^2,$ and
$r=t$ labels the point of contact between the interaction-potential operator
$g(r)$ and {\em its\/} upper-tangent approximation $a_2(t)+b_2(t)r.$
Consequently, our semirelativistic Hamiltonian $H$ in Eq.~(\ref{Eq:(sr)SSH})
satisfies the operator inequality$$H\equiv\beta k({\bf p}^2)+V(r)\le\bar
H(q,t)\equiv\beta\left[a_1(q)+b_1(q){\bf p}^2\right]+a_2(t)+b_2(t)r\ .$$The
eigenvalues $\bar E_{n\ell}(q,t)$ of the upper ``tangential Hamiltonian''
$\bar H(q,t)$ are given in terms~of the numbers $P_{n\ell}(1)$ listed in
Table~\ref{Tab:P(1)-numbers} by the exact formula\begin{equation}\bar
E_{n\ell}(q,t)=3\left[\frac{\beta
b_1(q)b_2^2(t)P_{n\ell}^2(1)}{4}\right]^{1/3}+\beta a_1(q)+a_2(t)\
.\label{Eq:comp-UEB}\end{equation}In order to find the best such upper bound,
we must minimize with respect to both $q$ and~$t.$ The critical equations
immediately yield the necessary
conditions\begin{equation}\frac{1}{2\beta}\frac{g'(\hat t)}{k'(\hat
q)}=\frac{\hat q^{3/2}}{P_{n\ell}(1)} =\frac{P_{n\ell}^2(1)}{\hat t^3}\
.\label{Eq:nec-cond}\end{equation}From the second of the above equalities we
deduce that, for given quantum numbers $n$ and~$\ell,$ the minimizing values
of the parameters $q$ and $t$ are constrained to the curve $\hat q\hat
t^2=P_{n\ell}^2(1)$~in the $(q,t)$ parameter plane. On this curve, and with
the coefficients (\ref{Eq:comp-coeff}) of our upper-tangent approximations
and the necessary (critical) conditions (\ref{Eq:nec-cond}), the minimum of
the expression on the right-hand side of the upper energy bound
(\ref{Eq:comp-UEB}) may be cast into a new, simpler~form:$$\min_{\hat
t>0}\left[\beta k\left(\frac{P_{n\ell}^2(1)}{\hat t^2}\right)+g(\hat
t)\right].$$By remembering the above definitions of $k({\bf p}^2)$ and
$g(r),$ this expression eventually becomes\begin{equation}
E_{n\ell}\le\min_{r>0}\left[\beta\sqrt{m^2+\frac{P^2}{r^2}}+V(r)\right],\quad
P\equiv P_{n\ell}(1)\ .\label{Eq:LUB}\end{equation}These upper bounds on the
entire discrete spectrum $\{E_{n\ell}\}$ are valid for any central potential
$V(r)$ which is a monotone increasing and concave function of the linear
potential $h(r)=r.$

\section{Variational upper bounds}\label{Sec:VarUB}The standard tool for the
derivation of rigorous upper bounds on the eigenvalues $E$ of some
self-adjoint operator $H$ is the Rayleigh--Ritz variational technique
\cite{Reed78,Thirring90}, which is~based~on the minimum--maximum~principle:
If the (discrete) eigenvalues $E_k,$ $k=0,1,2,\dots,$ of $H$~are ordered
according to $E_0\le E_1\le E_2\le\cdots,$ then the first $d$ of them are
bounded from above by the $d$ eigenvalues $\widehat E_k,$ $k=0,1,\dots,d-1,$
ordered according to $\widehat E_0\le\widehat E_1\le\cdots\le\widehat
E_{d-1},$~of that operator $\widehat H$ which is obtained by the restriction
of $H$ to a $d$-dimensional subspace $D_d$ of the domain of $H,$ that is,
$E_k\le\widehat E_k$ for all $k=0,1,\dots,d-1.$ Hence, for the
accuracy~of~the obtained results, an appropriate, ``reasonable'' definition
of this trial subspace $D_d$ is crucial. If this $d$-dimensional subspace
$D_d$ is spanned by a chosen set of $d$ linearly independent~basis vectors
$|\psi_k\rangle,$ $k = 0,1,\dots,d-1,$ the set of eigenvalues $\widehat E$
may immediately be determined~by diagonalizing the $d\times d$ matrix
$(\langle\psi_i|\widehat H|\psi_j\rangle),$ $i,j=0,1,\dots,d-1,$ that is, as
the $d$ roots~of~the characteristic equation$$\det(\langle\psi_i|\widehat
H|\psi_j\rangle-\widehat E\,\langle\psi_i|\psi_j\rangle)=0\ ,\quad
i,j=0,1,\dots,d-1\ .$$

The relativistic virial theorem derived in Ref.~\cite{Lucha89:RVT}---for a
very comprehensive review,~see Ref.~\cite{Lucha90:RVTs}---allows to define a
precise quantitative measure \cite{Lucha99Q,Lucha99A} for the quality of the
results found within the framework of variational techniques: according to
the analysis presented~in Refs.~\cite{Lucha99Q,Lucha99A}, for some generic
Hamiltonian operator $H$ consisting of a momentum-dependent kinetic-energy
term $T({\bf p})$ and a coordinate-dependent interaction-potential term
$V({\bf x}),$ i.e.,$$H=T({\bf p})+V({\bf x})\ ,$$the accuracy of trial states
$|\varphi\rangle$ which approximate the exact bound state under study~can~be
quantitatively estimated by the deviation from zero of the quantity (called
$\nu$ in Refs.~\cite{Lucha99Q,Lucha99A})$$Q\equiv1-\frac{\langle\varphi|{\bf
p}\cdot\frac{\partial}{\partial{\bf p}}T({\bf p})|\varphi\rangle}
{\langle\varphi|{\bf x}\cdot\frac{\partial} {\partial{\bf x}}V({\bf
x})|\varphi\rangle}\ (\equiv-\nu)\ .$$The main advantage of this measure for
the accuracy of approximate eigenstates $|\varphi\rangle$ is that it does not
require any information on the solutions of the investigated eigenvalue
problem other than the one provided by the variational approximation
technique itself.

In order to get a first idea of the position of the lowest energy level
$E_0,$ a one-dimensional trial~space is sufficient; a basis for the latter
might be any of the two-parameter (normalized) trial functions$$\phi({\bf
x})=\sqrt{\frac{n\alpha^{3/n}}{4\pi\Gamma(\frac{3}{n})}}
\exp\left(-\frac{\alpha}{2}r^n\right).$$Here the scale parameter $\alpha>0$
is varied. The parameter $n>0$ allows for later optimization. For the sake of
definiteness, let us consider an interaction potential $V(r)$ which is the
sum~of (attractive) pure power-law terms $a(q)\,{\rm sgn}(q)r^q$ with $q\ne0$
and a logarithmic term~$a(0)\ln r$:$$V(r)=\sum_{q\ne0}a(q)\,{\rm
sgn}(q)r^q+a(0)\ln r\ .$$We assume, of course, that the potential
coefficients $a(q)\ge0$ are not {\em all\/} zero. With the~help of Jensen's
inequality \cite{Jensen}, and by a tricky redefinition of the scale variable
$\alpha,$~the~variational upper bound on the ground-state eigenvalue $E_0$ of
the Hamiltonian (\ref{Eq:(sr)SSH}), resulting from this choice of $\phi({\bf
x})$ as basis vector of our one-dimensional trial space, may be cast into the
form\begin{equation}E_0\le\min_{r>0}\left[\beta\sqrt{m^2+\frac{1}{r^2}}+
\sum_{q\ne0}a(q)\,{\rm sgn}(q)({\cal P}(n,q)r)^q+a(0)\ln({\cal
P}(n,0)r)\right],\label{Eq:VUB}\end{equation}with the
abbreviations\begin{eqnarray*}{\cal
P}(n,q)&=&\frac{n}{2}\left(\frac{\Gamma(2+\frac{1}{n})}
{\Gamma(\frac{3}{n})}\right)^{1/2}
\left(\frac{\Gamma(\frac{q+3}{n})}{\Gamma(\frac{3}{n})}\right)^{1/q}\ ,\quad
q\neq 0\ ,\\[1ex]{\cal
P}(n,0)&=&\frac{n}{2}\left({\frac{\Gamma(2+\frac{1}{n})}
{\Gamma(\frac{3}{n})}}\right)^{1/2}
\exp\left(\frac{1}{n}\psi\left(\frac{3}{n}\right)\right),\end{eqnarray*}where
$\psi$ denotes the digamma function
$$\psi(z)\equiv\frac{1}{\Gamma(z)}\frac{{\rm d}\Gamma(z)}{{\rm d}z}\ .$$

\section{The ``Laguerre'' trial space}\label{Sec:LagTS}Unless the eigenstate
corresponding to some eigenvalue $E_k$ is already an element of the~trial
space $D_d,$ any upper bounds on the eigenvalues of an operator bounded from
below may~be improved by enlarging $D_d$ to higher $d,$ or by spanning $D_d$
by more sophisticated basis states.

For the class of spherically symmetric potentials $V(r)$ studied here, a very
popular~choice for the basis states which span the $d$-dimensional trial
space $D_d$ of the variational technique are ``Laguerre'' trial states, given
in their configuration-space representation by
\cite{LTS,Lucha97,Lucha98O,Lucha98D}\begin{equation}\psi_{k,\ell m}({\bf x})=
\sqrt{\frac{(2\mu)^{2\ell+2\rho+1}k!}{\Gamma(2\ell+2\rho+k+1)}}r^{\ell+\rho-1}
\exp(-\mu r)L_k^{(2\ell+2\rho)}(2\mu r){\cal Y}_{\ell m}(\Omega_{\bf x})\
.\label{eq:LagTF}\end{equation}Here $L_k^{(\gamma)}(x)$ are the generalized
Laguerre polynomials (for the parameter $\gamma$) \cite{Abramowitz},
defined~by the power series$$L_k^{(\gamma)}(x)=\sum_{t=0}^k(-1)^t
\left(\begin{array}{c}k+\gamma\\k-t\end{array}\right)\frac{x^t}{t!}$$and
orthonormalized, with the weight function $x^\gamma\exp(-x),$ according
to$$\int\limits_0^\infty{\rm d}x\,x^\gamma\exp(-x)L_k^{(\gamma)}(x)
L_{k'}^{(\gamma)}(x)=\frac{\Gamma(\gamma+k+1)}{k!}\delta_{kk'}\ ;$$${\cal
Y}_{\ell m}(\Omega)$ are the spherical harmonics for angular momentum $\ell$
and projection~$m,$ depending on the solid angle $\Omega$ and orthonormalized
according to$$\int{\rm d}\Omega\,{\cal Y}^\ast_{\ell m}(\Omega){\cal
Y}_{\ell'm'}(\Omega)=\delta_{\ell\ell'}\delta_{mm'}\ .$$The trial functions
(\ref{eq:LagTF}) involve two variational parameters, $\mu$ (with the
dimension of~mass) and $\rho$ (dimensionless), which, by the requirement of
normalizability of these functions, are subject to the constraints $\mu>0$
and $2\rho>-1.$

One of the main advantages of the choice (\ref{eq:LagTF}) for the variational
trial states is the easy availability of an analytic expression for their
momentum-space representation, obtained~by Fourier transformation of
$\psi_{k,\ell m}({\bf x})$:\begin{eqnarray*}\widetilde\psi_{k,\ell m}({\bf
p})&=&\sqrt{\frac{(2\mu)^{2\ell+2\rho+1}k!}{\Gamma(2\ell+2\rho+k+1)}}
\frac{(-{\rm i})^\ell|{\bf p}|^\ell}
{2^{\ell+1/2}\Gamma\left(\ell+\frac{3}{2}\right)}\sum_{t=0}^k\frac{(-1)^t}{t!}
\left(\begin{array}{c}k+2\ell+2\rho\\k-t\end{array}\right)\\[1ex]
&\times&\frac{\Gamma(2\ell+\rho+t+2)(2\mu)^t}{({\bf
p}^2+\mu^2)^{(2\ell+\rho+t+2)/2}}
F\left(\frac{2\ell+\rho+t+2}{2},-\frac{\rho+t}{2};\ell+\frac{3}{2};
\frac{{\bf p}^2}{{\bf p}^2+\mu^2}\right)\\[1ex]&\times&{\cal Y}_{\ell
m}(\Omega_{\bf p})\ ,\end{eqnarray*}with the hypergeometric series $F,$
defined by$$F(u,v;w;z)=\frac{\Gamma(w)}{\Gamma(u)\Gamma(v)}\sum_{n=0}^\infty
\frac{\Gamma(u+n)\Gamma(v+n)}{\Gamma(w+n)}\frac{z^n}{n!}\ .$$

The choice of the Laguerre states as the basis of the trial space $D_d$
allows to calculate~the matrix elements of (any linear combination of) pure
power-law potentials $V(r)=a(q)r^q$~and under certain circumstances (for
instance, for $\mu=m$ for {\em either\/} radial {\em or\/} orbital
excitations) the matrix elements of the kinetic term on purely analytical
grounds, and to arrive therefore at analytical expressions for the $d\times
d$ Hamiltonian matrix $(\langle\psi_i|\widehat H|\psi_j\rangle),$
$i,j=0,1,\dots,d-1.$ The explicit algebraic expressions of these matrix
elements may be found in Refs.~\cite{Lucha97,Lucha98O,Lucha98D}. Up to and
including a trial-space dimension $d=4,$ the Hamiltonian matrix
$(\langle\psi_i|\widehat H|\psi_j\rangle)$~may be diagonalized algebraically.
For the funnel potential (\ref{Eq:FP}), the case $d=1$ yields
the~bound$$E_0\le\left(\frac{64\beta}{15\pi}-c_1\right)m+\frac{3c_2}{2m}\ .$$

\section{The ``local-energy'' theorem}\label{Sec:LET}Some information on {\em
both\/} upper {\em and\/} lower bounds on the isolated eigenvalues
$E_{n\ell}$ of some Hamiltonian $H$ may be gained with the help of the
so-called ``local-energy'' theorem~\cite{LET,Thirring90}. Unfortunately, the
proof of this fundamental criterion makes use of the ``nodal theorem''~for
the eigenstates of $H.$ As a consequence of this, for Hamiltonians $H$
defined in more than~one dimension the local-energy theorem can be applied
only to the ground state of the spectrum. It has been applied to the
(three-dimensional) ``relativistic Coulomb problem'' in Ref.~\cite{Raynal94}.

Upper bounds on the energy levels of Hamiltonians bounded from below can be
obtained with considerably more efficiency by the variational technique
described in Sec.~\ref{Sec:VarUB}. Therefore we employ here the local-energy
theorem for the derivation of {\em lower\/} bounds on the spectrum of $H.$ In
order to formulate the local-energy theorem in momentum-space
representation,~we introduce the ``local energy''$${\cal E}({\bf p})\equiv
\beta\sqrt{m^2+{\bf p}^2}+\frac{\displaystyle\int{\rm d}^3q\,\widetilde
V({\bf p}-{\bf q})\phi({\bf q})} {\phi({\bf p})}\ .$$Here, $\widetilde V({\bf
p})$ is the Fourier transform of the interaction potential $V({\bf x})$ under
consideration:$$\widetilde V({\bf p})\equiv\frac{1}{(2\pi)^3}\int{\rm
d}^3x\exp(-{\rm i}{\bf p}\cdot{\bf x})V({\bf x})\ .$$$\phi({\bf p})$ denotes
a suitably chosen, positive trial function: $\phi({\bf p})>0.$ The
``lower-bound part''~of the local-energy theorem then states that the
lowest-lying eigenvalue $E_0$ of the Hamiltonian $H$ is bounded from below by
the infimum of ${\cal E}({\bf p})$ with respect to the momentum
variable~${\bf p}$:\begin{equation}E_0\ge\inf_{{\bf p}}{\cal E}({\bf p})\
.\label{Eq:LET}\end{equation}

The application of the local-energy theorem to {\em confining\/} potentials
demands special~care for the following reason: the Fourier transform of any
confining potential is not well-defined; the potential has to be regarded as
a {\em distribution\/} and an appropriate regularization must~be applied. For
instance, the linear potential $V_{\rm L}(r)=\lambda r,$ where $\lambda$ is a
coupling parameter with mass dimension 2, may be regularized by the
exponential $\exp(-\varepsilon r)$, $\varepsilon\ge0,$ or by writing~it as
the derivative of this exponential or as the second derivative of a
Yukawa-type function:$$V_{\rm L}(r)=\lambda
r=\lambda\lim_{\varepsilon\downarrow0}r\exp(-\varepsilon
r)=-\lambda\lim_{\varepsilon\downarrow0}\frac{\partial}{\partial\varepsilon}
\exp(-\varepsilon r)=\lambda\lim_{\varepsilon\downarrow0}
\frac{\partial^2}{\partial\varepsilon^2}\frac{\exp(-\varepsilon r)}{r}\ .$$

\newpage\noindent With these regularizations, the Fourier transform
$\widetilde V_{\rm L}({\bf p})$ of the linear potential $V_{\rm L}(r)$
reads$$\widetilde V_{\rm L}({\bf p})=\widetilde V_{\rm L}(|{\bf
p}|)=-\frac{\lambda}{\pi^2}\lim_{\varepsilon\downarrow0}\frac{{\bf
p}^2-3\varepsilon^2}{({\bf p}^2+\varepsilon^2)^3}\ .$$Note that (as a
consistency check) the integral of the Fourier transform $\widetilde V_{\rm
L}({\bf p})$ has to vanish: $\int{\rm d}^3p\,\widetilde V_{\rm L}({\bf
p})=0.$ This holds for every power-law potential $V(r)=a_nr^n$ with
exponent~$n>0.$ The singularity structure of $\widetilde V_{\rm L}({\bf p})$
becomes manifest when writing the derivative in $V_{\rm L}(r)$~as a
difference quotient,$$V_{\rm L}(r)=\lambda r
=-\lambda\lim_{\varepsilon\downarrow0}\frac{\exp(-\varepsilon
r)-1}{\varepsilon}\ ,$$which yields$$\widetilde V_{\rm L}({\bf p})=\widetilde
V_{\rm L}(|{\bf p}|)=\lambda\lim_{\varepsilon\downarrow0}
\left[\frac{1}{\varepsilon}\delta^{(3)}({\bf p})-\frac{1}{\pi^2({\bf
p}^2+\varepsilon^2)^2}\right].$$Thus the Fourier transform $\widetilde V_{\rm
L}({\bf p})$ of the linear potential consists of a (negative) regular~part
and a singular part, which is a distribution localized at the origin ${\bf
p}={\bf 0}.$ However, a detailed inspection shows that for the case of the
funnel potential the momentum-space~ground-state eigenfunction is indeed
positive. This fact justifies our use of a positive trial function~$\phi({\bf
p}).$

The simplest normalizable trial function which comes to one's mind is the
exponential:$$\phi(r)=\sqrt{\frac{\mu^3}{\pi}}\exp(-\mu r)\ .$$This choice
for the ground-state trial function $\phi(r)$ corresponds to the case
$k=\ell=m=0$ and $\rho=1$ of the Laguerre basis functions $\psi_{k,\ell
m}({\bf x})$ introduced in Sec.~\ref{Sec:LagTS}. It is~straightforward to
calculate the Fourier transform of $\phi(r),$ or to extract it from the
general expression given explicitly in Sec.~\ref{Sec:LagTS}:$$\phi({\bf
p})=\phi(|{\bf p}|)=\frac{\sqrt{2\mu^3}}{\pi|{\bf p}|({\bf p}^2+\mu^2)}
\sin\left(2\arctan\frac{|{\bf p}|}{\mu}\right)=\frac{2\sqrt{2\mu^5}}{\pi({\bf
p}^2+\mu^2)^2}\ .$$In order to illustrate the application of the local-energy
theorem, let us consider the example of a potential $V(r)$ which is a sum of
pure power-law terms:$$V(r)=\sum_qa(q)r^q\ .$$For this power-law potential
and the ground-state trial function $\phi(r),$ it is straightforward~to
obtain the corresponding local energy ${\cal E}({\bf p})$ by Fourier
transformation of the product $r^q\phi(r)$: $${\cal E}({\bf
p})=\beta\sqrt{m^2+{\bf p}^2}+
\sum_q\frac{a(q)\Gamma(q+2)}{2\mu|{\bf p}|({\bf p}^2+\mu^2)^{q/2-1}}
\sin\left[(q+2)\arctan\frac{|{\bf p}|}{\mu}\right].$$The variational
parameter $\mu$ entering into our ground-state trial function $\phi(r)$
either may~be given a chosen fixed value, or it may be adjusted in order to
maximize the lower bound~(\ref{Eq:LET}).

\section{Sums of distinct potential terms}\label{Sec:SLB}For the class of
potentials $V(r)$ which consist of two or more distinct terms $V^{(i)}(r)\ne
V^{(j)}(r)$ for $i\ne j,$ where, for every single component problem defined
by the ``one-term'' Hamiltonian $H_i\equiv\beta\sqrt{m^2+{\bf
p}^2}+V^{(i)}(r),$ information about the bottom of its spectrum is available,
these individual pieces of information may be combined to form a lower bound
on the spectrum~of the Hamiltonian (\ref{Eq:(sr)SSH}) with such a sum
potential $V(r).$ The precise requirements for this ``sum approximation'' to
work are the following:\begin{enumerate}\item The interaction potential
$V(r)$ entering in the semirelativistic Hamiltonian of Eq.~(\ref{Eq:(sr)SSH})
is a sum of (more than one) different components $V^{(i)}(r)=c_ih^{(i)}(r)$:
$$V(r)=\sum_ic_ih^{(i)}(r)\ .$$\item Every component problem
$H_i\equiv\beta\sqrt{m^2+{\bf p}^2}+c_ih^{(i)}(r)$ supports, for
sufficiently~large values of its coupling parameter $c_i,$ a discrete
eigenvalue at the bottom of its spectrum.\end{enumerate}The sum approximation
has already been studied for the case of Schr\"odinger Hamiltonians
(involving nonrelativistic kinetic energies) in Refs.~\cite{Hall83,NRSA}. The
analysis of the corresponding semirelativistic problem has been presented in
full generality in Ref.~\cite{Lucha02-Sum}. Therefore, here~we summarize only
the result for the case in which the potential $V(r)$ is a sum of
attractive~pure power-law terms:\begin{equation}V(r)=\sum_{q\ne0}a(q)\,{\rm
sgn}(q)r^q\ ,\label{Eq:SPPP}\end{equation}where the coupling parameters
$a(q)$ are non-negative and not all zero. For this potential, a lower bound
to the bottom of the spectrum of the semirelativistic Hamiltonian
(\ref{Eq:(sr)SSH}), which~is, by assumption, the lowest eigenvalue $E_0$ of
$H,$ is given by the expression
\begin{equation}E_0\ge\min_{r>0}\left[\beta\sqrt{m^2+\frac{1}{r^2}}+
\sum_{q\ne0}a(q)\,{\rm sgn}(q)(P(q)r)^q\right],\label{Eq:SLB}\end{equation}
{\em if}, for the particular potential $V(r)$ under consideration, there
exist (or there can be found) $P$ numbers $P(q)$ such that, whenever there is
only one term present in the potential (\ref{Eq:SPPP}), the right-hand side
of the inequality (\ref{Eq:SLB}) yields either the exact ground-state energy
eigenvalue of the corresponding ``single-term'' problem or a lower bound to
it.

Consequently, when applying our general sum lower bound (\ref{Eq:SLB}) to a
given sum potential $V(r),$ the main task is to derive appropriate $P$
numbers for all components in the sum~(\ref{Eq:SPPP}). Let us illustrate this
by considering the two examples relevant for the funnel potential
(\ref{Eq:FP}).\begin{itemize}\item For a Coulomb component potential
$h^{(i)}(r)=-1/r,$ corresponding to the case $q=-1$ in Eq.~(\ref{Eq:SPPP}),
we may take advantage from the explicit results presented in
Sec.~\ref{Sec:CLB}. By~the change of variables $r\to Pr,$ the Coulomb lower
bound (\ref{Eq:SSE-LB}) may be cast into the form$$E_0\ge\min_{r>0}
\left[\beta\sqrt{m^2+\frac{1}{r^2}}+V(Pr)\right],\quad v<\beta v_P\
.$$Specifying this general result to the Coulomb potential $V(r)=-v/r$ yields
the bound
$$E_0\ge\min_{r>0}\left[\beta\sqrt{m^2+\frac{1}{r^2}}-\frac{v}{Pr}\right],\quad
v<\beta v_P\ .$$This bound is, of course, nothing else but, for $P=2/\pi,$
the Herbst lower bound~(\ref{Eq:RCP-HLB})~or, for arbitrary $P,$ the
Herbst-like lower bound (\ref{Eq:RCP-NLB}) to the relativistic Coulomb
problem. Consequently, for the Coulomb component term we may identify the $P$
number $P(-1)$ required for the lower bound (\ref{Eq:SLB}) with the
lower-bound parameter $P$ defined in Sec.~\ref{Sec:CLB}: $P(-1)=P.$\item For
a linear component potential $h^{(i)}(r)=r,$ that is, for the case $q=1$ in
Eq.~(\ref{Eq:SPPP}),~(a possible value of) the $P$ number $P(1)$ required for
the lower bound (\ref{Eq:SLB}) turns out to~be related to the ground-state
eigenvalue $e_0(v)$ of the Hamiltonian operator $\sqrt{{\bf
p}^2}+vr$~by$$e_0(v)=2\sqrt{vP(1)}\ ;$$a numerical determination of this
lowest energy eigenvalue $e_0(v)$ yields $P(1)=1.2457.$\end{itemize}

\section{Comparison: The case of the funnel potential}\label{Sec:Comp}In
order to demonstrate the power of the different energy bounds derived so far,
let us apply the above results to the very illustrative example of the
Coulomb-plus-linear potential~(\ref{Eq:FP}). By factorizing off an overall
coupling constant $0<v\le1,$ we write this potential in the~form
\begin{equation}V(r)=v\left(-\frac{a}{r}+br\right).\label{Eq:CLP}\end{equation}
For the coupling parameters entering in this funnel potential $V(r),$
different notations have been used in Eqs.~(\ref{Eq:FP}), (\ref{Eq:SPPP}),
and (\ref{Eq:CLP}). Evidently, the three sets of potential coefficients~have
to be identified according to $a(-1)\equiv c_1\equiv av>0$ and $a(1)\equiv
c_2\equiv bv>0.$

\begin{figure}[ht]\vspace*{-2cm}\begin{center}
\psfig{figure=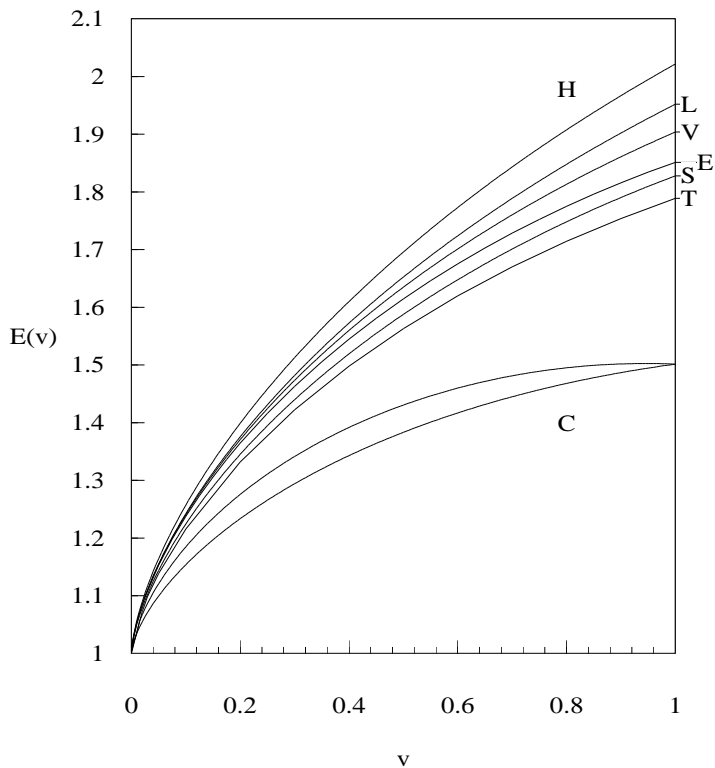,scale=.6243}\vspace*{-1cm}\caption{Comparison of
bounds to the ground-state [$(n,\ell)=(1,0)$] energy eigenvalue of the
semirelativistic Hamiltonian $H=\beta\sqrt{m^2+{\bf p}^2}+V(r)$ with a
Coulomb-plus-linear~potential $V(r)=v(-a/r+br),$ for $a=0.2,$ $b=0.5,$
$m=\beta=1.$ The bounds are plotted as functions $E(v)$ of the overall
coupling parameter $v$ in $V(r).$ The energy bounds compared~here include
(from top to bottom) the harmonic-oscillator upper bound (H) of
Eq.~(\ref{Eq:SSE-HOUB}), the linear upper bound (L) of Eq.~(\ref{Eq:LUB}),
the ``one-dimensional'' variational upper bound (V) of Eq.~(\ref{Eq:VUB})
with $n=1.74,$ the variational upper bound (E) based on a 25-dimensional
Laguerre trial space, which should come rather close to the exact eigenvalue,
the sum lower bound (S) of Eq.~(\ref{Eq:SLB}) with $P$ numbers
$P(-1)=0.72811$ and $P(1)=1.2457,$ the lower bound (T) provided by~the
local-energy theorem (\ref{Eq:LET}), and two Coulomb lower bounds (C)
according to Eq.~(\ref{Eq:SSE-LB}),~where with respect to the parameter $P$
the lower curve corresponds to the fixed value $P=0.72811$ whereas the upper
curve corresponds to the optimized value of $P$ calculated from
Eq.~(\ref{Eq:P(m)}).}\label{Fig:SSE-bounds}\end{center}\end{figure}

Fig.~\ref{Fig:SSE-bounds} compares the various energy bounds derived or
investigated in Secs.~\ref{Sec:CLB} through~\ref{Sec:SLB} for the
ground-state energy eigenvalue $E_0$ of the Hamiltonian (\ref{Eq:(sr)SSH})
with the funnel potential (\ref{Eq:CLP}) as a function of the overall
coupling parameter $v.$ These energy bounds necessarily~form a bunch of
concave curves $E(v),$ all starting at the (free-energy) value
$E_0(v=0)=\beta m=1.$

The Coulomb lower bound (\ref{Eq:SSE-LB}) may be applied to the funnel
potential either with~a~fixed value of the lower-bound parameter $P$ or with
the optimized $P(v)$ computed from Eq.~(\ref{Eq:P(m)}). For ``fixed-$P$''
bounds, the maximum value of $P$ allowed by the Coulomb coupling constant
constraint discussed in Sec.~\ref{Sec:CLB} is found, for given values of the
kinetic-term parameters~$\beta,$~$m$ and potential coefficients $c_1,$ $c_2,$
as solution of that relation which is obtained when equating the left-hand
and right-hand sides of the inequality (\ref{Eq:CCCC}) with the critical
coupling $v_P$ given by the definition (\ref{Eq:BV}), that is, by solving,
for the {\em maximum\/} values of $c_1$ and $c_2$ occuring~here, the
constraint~(\ref{Eq:P(m)}):$$c_1+\frac{P^4}{1-P^2}\frac{c_2}{m^2}=\beta
P\sqrt{1-P^2}\ .$$For a single, unit-mass particle, i.e., $\beta=1$ and
$m=1,$ and the values of the funnel-potential coupling parameters $a$ and $b$
used in Fig.~\ref{Fig:SSE-bounds}, $a=0.2$ and $b=0.5,$ this yields
$P=0.728112.$

We try to approach the state we are interested in as closely as possible by
the application of the Rayleigh--Ritz variational technique brief\/ly
sketched in Sec.~\ref{Sec:VarUB}, with the trial space~$D_d$ spanned by the
Laguerre basis defined in Sec.~\ref{Sec:LagTS}. By fixing the variational
parameter $\mu$ to~the value $\mu=m,$ we are able to obtain the matrix
elements of the Hamiltonian (\ref{Eq:(sr)SSH}) analytically, avoiding thereby
the necessity to use a numerical integration procedure for~the evaluation~of
these matrix elements; for definiteness, we fix the variational parameter
$\rho$ to the value~$\rho=1.$

\begin{figure}[ht]\vspace*{-0.395cm}\begin{center}
\psfig{figure=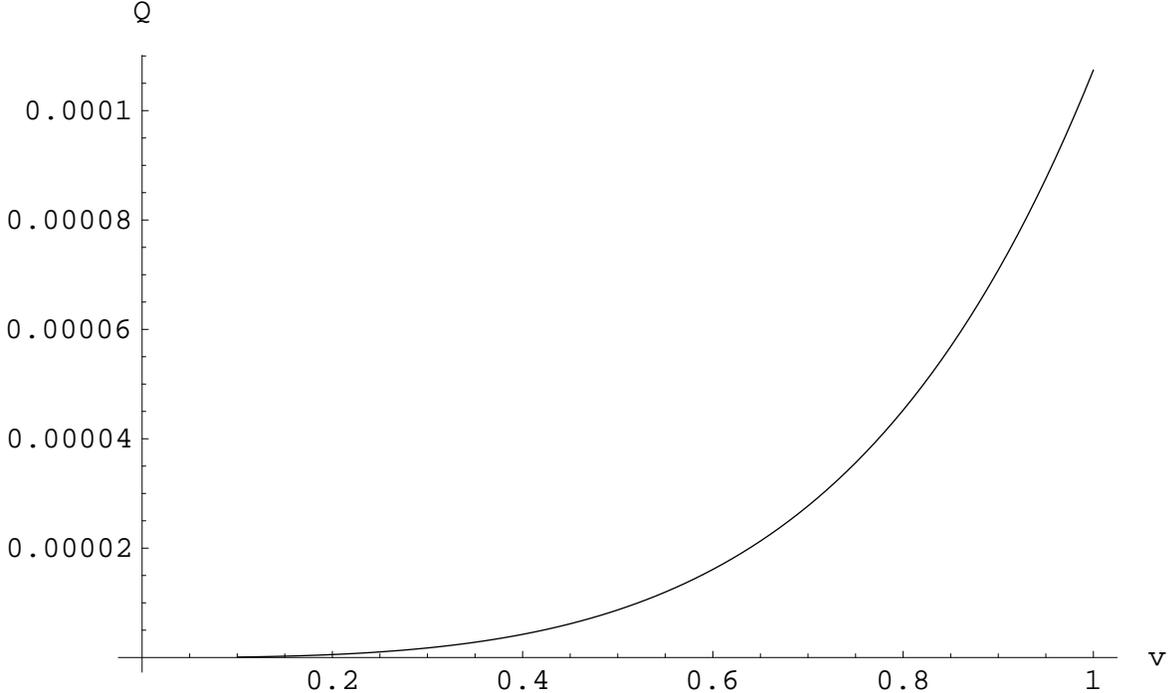,scale=1.1416}\caption{Measure $Q$ of the accuracy
of the (variational) approximation of the ground-state eigenfunction~of the
semirelativistic Hamiltonian of Fig.~\ref{Fig:SSE-bounds} [that is,
$H=\beta\sqrt{m^2+{\bf p}^2}+V(r),$ where $V(r)$ is the Coulomb-plus-linear
potential $V(r)=v(-a/r+br)$ with $a=0.2,$ $b=0.5,$ $m=\beta=1$] by a
superposition of 25 Laguerre basis functions (\ref{eq:LagTF}) with $\mu=m$
and $\rho=1.$}\label{Fig:accuracy}\end{center}\end{figure}

The accuracy of such an approximation to an eigenstate may be estimated with
the~help of the ``virial-theorem-inspired'' quality measure $Q$ discussed in
Sec.~\ref{Sec:VarUB}. Fig.~\ref{Fig:accuracy} shows, for~the trial-space
dimension $d=25$ chosen for the variational bound (E) in
Fig.~\ref{Fig:SSE-bounds}, the dependence of the accuracy $Q$ on the overall
coupling $v.$ We obtain $Q=|Q|\le1.1\times 10^{-4}$ over the~whole parameter
range $0<v\le1.$ Such high accuracy should be sufficient for the present
purpose.

\begin{figure}[ht]\begin{center}\psfig{figure=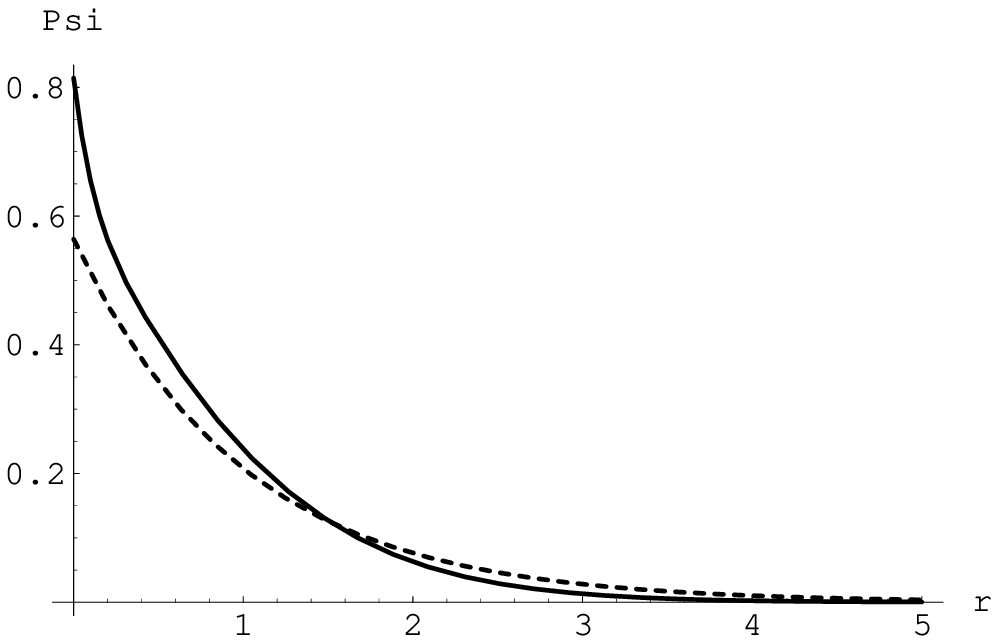,scale=1.1658}
\\[1ex](a)\\[1ex]\psfig{figure=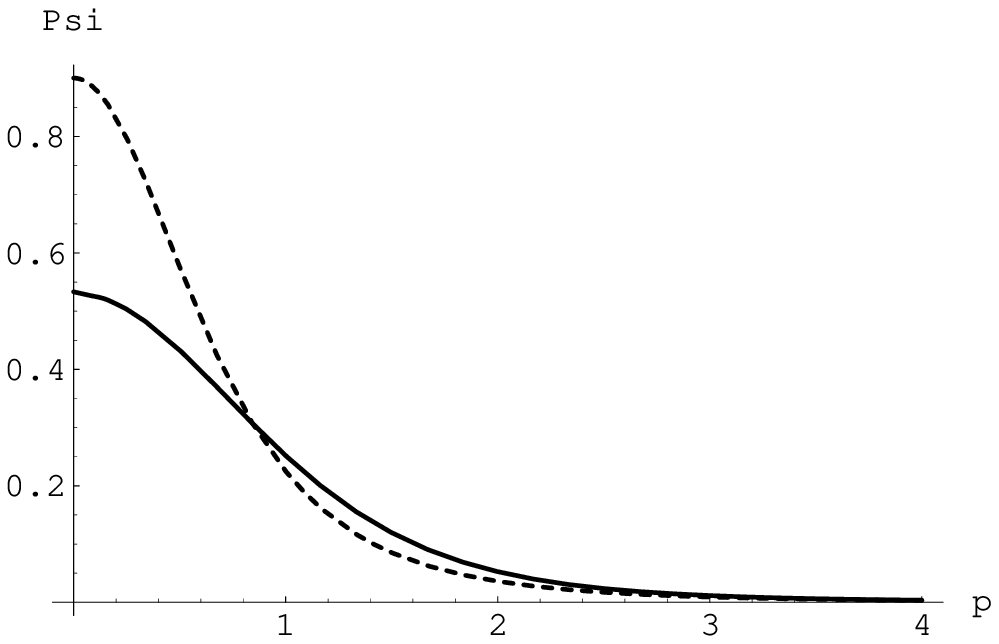,scale=1.1658}\\[1ex](b)
\caption{Variational approximation of the lowest-lying eigenstate in
configuration space~(a) and momentum space (b) of the semirelativistic
Hamiltonian (\ref{Eq:(sr)SSH}), $H=\beta\sqrt{m^2+{\bf p}^2}+V(r),$ with a
Coulomb-plus-linear potential $V(r)=-a/r+br,$ for $a=0.2,$ $b=0.5$ and
$m=\beta=1,$ by a superposition of Laguerre basis functions (\ref{eq:LagTF})
with the variational parameters fixed~to $\mu=m$ and $\rho=1,$ for the
trial-space dimensions $d=25$ (full lines) and $d=1$ (dashed~lines).}
\label{Fig:Laguerre-approx}\end{center}\end{figure}

For the sake of completeness, we plot, in Fig.~\ref{Fig:Laguerre-approx}, the
behaviour of the ground state of~the Hamiltonian (\ref{Eq:(sr)SSH}) with the
Coulomb-plus-linear potential (\ref{Eq:CLP}), for the maximum value $v=1$ of
the overall coupling parameter $v,$ determined by the variational technique
of Sec.~\ref{Sec:VarUB}, with the Laguerre basis of Sec.~\ref{Sec:LagTS}, in
both configuration and momentum space. As expected, the ground-state
momentum-space eigenfunction possesses no node, which~is in~agreement with
our assumption of a positive momentum-space trial function $\phi({\bf p})$
(which is also depicted,~as the case $d=1,$ in
Fig.~\ref{Fig:Laguerre-approx}) in the derivation of the local-energy lower
energy bound in Sec.~\ref{Sec:LET}.

\section{Summary and conclusions}This review of the discrete spectra of
semirelativistic Hamiltonians $H$ of the form (\ref{Eq:(sr)SSH})~should be
considered as a compilation of rigorous (semi-)analytical results for both
upper and~lower bounds on the discrete eigenvalues $E_{n\ell}$ of $H,$
presented in a way suitable for immediate~use: the algebraic structure and
the convexity properties with respect to suitable basis potentials $h(r)$~of
the interaction potential $V(r)$ one wishes to use in one's semirelativistic
Hamiltonian (\ref{Eq:(sr)SSH}) unambiguously determine which of all the
energy bounds discussed here actually apply. With the explicit analytical
expressions for these bounds on the discrete eigenvalues at~one's disposal,
one is left with the (numerical) optimization with respect to a single real
variable. Clearly, within the respective limitations, the Rayleigh--Ritz
variational technique of Secs.~\ref{Sec:VarUB} and \ref{Sec:LagTS} as well as
the local-energy theorem of Sec.~\ref{Sec:LET} apply to {\em arbitrary\/}
reasonable potentials.

For the particular example considered in Sec.~\ref{Sec:Comp}, viz., the
Coulomb-plus-linear potential (\ref{Eq:CLP}), the sum lower bound turns out
to be somewhat better than the lower bound provided by the local-energy
theorem. Clearly, the sum lower bound cannot always be better than the
local-energy bound because one could, in principle, choose in the
local-energy theorem~(\ref{Eq:LET}) the exact eigenfunction for the trial
function $\phi$ and thus recover the exact energy eigenvalue.

In conclusion, let us stress that all but two of the various energy bounds we
have studied in this work may eventually be expressed by essentially the same
general semi-classical~form. We can view this in the following way: Since the
square of the momentum~${\bf p}^2$ scales like~$1/r^2,$ we expect
that$$\langle{\bf p}^2\rangle=\frac{P^2}{r^2}\ ,$$where $P$ is a suitable
coefficient; thus the balance between the kinetic and potential energies
required by the minimum--maximum principle might lead to a search for the
minimum of~an energy expression of the form$$\sqrt{m^2+\frac{P^2}{r^2}}+V(r)\
.$$What makes this heuristic argument interesting is the fact that definite
values of $P$ can be prescribed so that this minimum provides {\it a priori}
a bound on the (unknown) exact energy. In the case where $V(r)$ is a sum of
terms, a trivial change of variables is used to move~the~$P$ parameters
inside the potential $V(r),$ and then optimal distinct parameters~can be
used~for each potential term.

The lower bound by the sum approximation incorporates our best known lower
bounds for the sub-problems containing only a single potential term; the
final result is the optimized mixture of these contributions. The lower bound
by the local-energy theorem (\ref{Eq:LET}) could, in principle, achieve
arbitrarily high accuracy. The difficulty here, as with variational methods,
is to find a suitable trial function. To the extent that we have searched we
did not manage to find a better lower bound with the help of the local-energy
theorem (\ref{Eq:LET}) than that~provided immediately by the sum
approximation (\ref{Eq:SLB}).

Our most accurate results overall were upper bounds obtained by using a
25-dimensional trial space spanned by Laguerre functions. We estimate the
error for these results to be less than $0.01\%$ over the total range of
potential parameters studied; such accurate estimates~are invaluable for
estimating the effectiveness of the large variety of alternative simpler
energy bounds which we have reviewed in this article.

\end{document}